%


\def\typeout#1{\immediate\write16{#1}}

\typeout{  >>>>>>>>>>>>>  LOADING MG7TEX PLAIN MACROS  <<<<<<<<<<<<<<<}
\typeout{  >>>>>>>>>>>>>  but you should be using LATEX!!  <<<<<<<<<<<<<<<}

\def\LaTeX{{\rm L\kern-.36em\raise.3ex\hbox{\sc a}\kern-.15em
    T\kern-.1667em\lower.7ex\hbox{E}\kern-.125emX}}

\def\newline{\hfil\break\typeout{-->Forced linebreak}}


\newskip\narrowskipamount \narrowskipamount=\parindent
\def\narrower{\advance\leftskip by \narrowskipamount
  \advance\rightskip by \narrowskipamount}


\def\pagesetup{
\hsize 6.0in
\vsize 8.5in
\advance \hoffset by .25in                 
\raggedbottom\parindent=20pt
\baselineskip=14pt}

\pagesetup



\font\twelverm=cmr10 scaled 1200
\font\twelvei=cmmi10 scaled 1200
\font\twelvesy=cmsy10 scaled 1200
\font\twelvebf=cmbx10 scaled 1200 
\font\twelvett=cmtt10 scaled 1200
\font\twelveit=cmti10 scaled 1200
\font\twelvesl=cmsl10 scaled 1200
\font\twelvesc=cmcsc10 scaled 1200
\font\twelvess=cmss10 scaled 1200 
\font\twelvessi=cmssi10 scaled 1200
\font\twelvebmit=cmmib10 scaled 1200
     
\font\tensc=cmcsc10
\font\tenss=cmss10
\font\tenssi=cmssi10
\font\tenbmit=cmmib10
     
\font\ninerm=cmr9        \font\sixrm=cmr6
\font\ninei=cmmi9    \font\eighti=cmmi8    \font\sixi=cmmi6
\font\ninesy=cmsy9   \font\eightsy=cmsy8   \font\sixsy=cmsy6
\font\ninebf=cmbx9      \font\sixbf=cmbx6
\font\ninett=cmtt9   
\font\nineit=cmti9      
\font\ninesl=cmsl9   
\font\niness=cmss9   
\font\ninessi=cmssi9 
     
     
\skewchar\twelvei='177    \skewchar\ninei='177
\skewchar\eighti='177     \skewchar\sixi='177
\skewchar\twelvesy='60    \skewchar\ninesy='60
\skewchar\eightsy='60     \skewchar\sixi='60
     
     
\newskip\ttglue
     
\def\twelvept{\def\rm{\fam0\twelverm}
   \textfont0=\twelverm \scriptfont0=\ninerm \scriptscriptfont0=\sevenrm%
   \textfont1=\twelvei  \scriptfont1=\ninei  \scriptscriptfont1=\seveni%
   \textfont2=\twelvesy \scriptfont2=\ninesy \scriptscriptfont2=\sevensy%
   \textfont3=\tenex    \scriptfont3=\tenex  \scriptscriptfont3=\tenex%
   \textfont\itfam=\twelveit   \def\it{\fam\itfam\twelveit}%
   \textfont\slfam=\twelvesl   \def\sl{\fam\slfam\twelvesl}%
   \textfont\ttfam=\twelvett   \def\tt{\fam\ttfam\twelvett}%
   \textfont\bffam=\twelvebf   \scriptfont\bffam=\ninebf%
      \scriptscriptfont\bffam=\sevenbf   \def\bf{\fam\bffam\twelvebf}%
   \def\oldstyle{\fam1 \twelvei}%
   \tt \ttglue=.5em plus.25em minus.15em%
   \normalbaselineskip=14pt%
   \setbox\strutbox=\hbox{\vrule height9pt depth4pt width0pt}%
   \let\sc=\twelvesc   \let\big=\tenbig%
   \let\ss=\twelvess   \let\ssi=\twelvessi   \let\bmit=\twelvebmit%
   \normalbaselines\rm}

\def\tenpt{\def\rm{\fam0\tenrm}
   \textfont0=\tenrm \scriptfont0=\sevenrm \scriptscriptfont0=\fiverm%
   \textfont1=\teni  \scriptfont1=\seveni  \scriptscriptfont1=\fivei%
   \textfont2=\tensy \scriptfont2=\sevensy \scriptscriptfont2=\fivesy%
   \textfont3=\tenex \scriptfont3=\tenex   \scriptscriptfont3=\tenex%
   \textfont\itfam=\tenit   \def\it{\fam\itfam\tenit}%
   \textfont\slfam=\tensl   \def\sl{\fam\slfam\tensl}%
   \textfont\ttfam=\tentt   \def\tt{\fam\ttfam\tentt}%
   \textfont\bffam=\tenbf   \scriptfont\bffam=\sevenbf%
      \scriptscriptfont\bffam=\fivebf   \def\bf{\fam\bffam\tenbf}%
   \def\oldstyle{\fam1 \teni}%
   \tt \ttglue=.5em plus.25em minus.15em%
   \normalbaselineskip=12pt%
   \setbox\strutbox=\hbox{\vrule height8.5pt depth3.5pt width0pt}%
   \let\sc=\tensc   \let\big=\tenbig%
   \let\ss=\tenss   \let\ssi=\tenssi   \let\bmit=\tenbmit%
   \normalbaselines\rm}
     
\def\ninept{\def\rm{\fam0\ninerm}
   \textfont0=\ninerm \scriptfont0=\sixrm \scriptscriptfont0=\fiverm%
   \textfont1=\ninei  \scriptfont1=\sixi  \scriptscriptfont1=\fivei%
   \textfont2=\ninesy \scriptfont2=\sixsy \scriptscriptfont2=\fivesy%
   \textfont3=\tenex  \scriptfont3=\tenex \scriptscriptfont3=\tenex%
   \textfont\itfam=\nineit   \def\it{\fam\itfam\nineit}%
   \textfont\slfam=\ninesl   \def\sl{\fam\slfam\ninesl}%
   \textfont\ttfam=\ninett   \def\tt{\fam\ttfam\ninett}%
   \textfont\bffam=\ninebf   \scriptfont\bffam=\sixbf%
      \scriptscriptfont\bffam=\fivebf   \def\bf{\fam\bffam\ninebf}%
   \def\oldstyle{\fam1 \ninei}%
   \tt \ttglue=.5em plus.25em minus.15em%
   \normalbaselineskip=10pt%
   \setbox\strutbox=\hbox{\vrule height8pt depth3pt width0pt}%
   \let\sc=\sevenrm   \let\big=\ninebig%
   \let\ss=\niness    \let\ssi=\ninessi%
   \normalbaselines\rm}

   \twelvept

%


\def\Footnote#1#2{\footnote{#1}{\ninerm #2\hfil}}

\typeout{ >>  MG7tex note: Remember to use the new capitalized 
              footnote command}

\typeout{ >>  MG7tex note: use  * , backslash "dag" , backslash "ddag"  
              for argument of titlepage 
              footnote superscript symbol and lowercase Latin letters 
              for other footnotes}

  \typeout{ >> MG7tex note: If you are looking to squeeze out white space,
               you might consider reducing the .8cm length used in the title
               material to .5cm. I have done this.}

\def\\{\par}

\long\def\beginCenter#1\endCenter{{\parskip=0pt
   \leftskip=0pt plus 1fil \rightskip = 0pt plus 1fil \parfillskip = 0pt 
   #1\par
   }}

\def\Title#1\endTitle{
     \baselineskip=16pt {\bf #1}
     \vskip.5cm}
 
  \typeout{ >> MG7tex note: Note that long titles must be broken by hand
               and should be all uppercase.}
\def\beginAuthors#1\endAuthors{\tenpt\it #1
                \vglue .2cm  
                \rm}

\def\Author#1#2\endAuthor{%
   {\rm #1}\\\vskip1pt #2\\\vskip.3cm}

  \typeout{ >> MG7tex note: Note that author names should be all uppercase.}


  \typeout{ >> MG7tex note: Note that no blanklines may appear 
               in the titlepage before the endAuthors command 
               (center environment in use).}

\def\Abstract#1{{\tenpt
            \centerline{ABSTRACT}
              \centerline{\vtop{\hsize=30pc 
                 \rm\noindent
                 #1}}
             } 
          \twelvept}


\typeout{ >> MG7tex note: Warning! These sectioning commands are not 
intelligent enough to avoid bad breaks between the section title and text.}



\newdimen\sechang  \setbox9=\hbox{\bf 0. \strut}
    \sechang=0pt \advance\sechang by \wd9
\newdimen\subsechang  \setbox9=\hbox{\bf 0.0. \strut}
    \subsechang=0pt \advance\subsechang by \wd9
 
\def\section#1{
\vskip .5truecm plus -0.5ex minus -0.1ex
{\noindent\hangindent \sechang \bf #1}
\medskip
}

\def\subsection#1{
\ifdim\lastskip=\medskipamount \vskip2pt \else\bigskip\fi
{\noindent\hangindent 20pt \it #1}
\smallskip
}

\def\subsubsection#1{
\ifdim\lastskip=\smallskipamount \else\smallskip\fi
{\noindent\hangindent 40pt \rm #1}  
\smallskip
}


\typeout{  >> MG7tex note: Use the cite command with numbered reference
              arguments for superscript references in the text}

\def\cite#1{${}^{#1}$}


\chardef\bslash=`\\    


\typeout{>>>>>>>>>>> End MG7 Plain TeX proceedings macros <<<<<<<<<<<}


\input epsf
\overfullrule=0pt
\def\list#1#2{\noindent\hangindent 5em\hangafter1\hbox to 
5em{\hfil#1\quad}#2}
\def\cli{\centerline}
\def\ora{\overrightarrow}
\def\n{\noindent}

\def\zero{\footline={\ifnum\pageno>0\hss\folio\hss\fi}\pageno=0}
\def\exam#1#2{\dimen1=2.25em
                      \noindent\hangindent=\dimen1\hangafter1
                      \hbox to\dimen1{#1\hfil~~}#2} 
\def\exama#1#2#3{\dimen1=2.25em \dimen2=4.5em
                      \noindent\hangindent=\dimen2\hangafter1
                      \hbox to\dimen1{#1\hfil~~}\hbox to\dimen1{#2\hfil~~}#3} 
\def\ctablel#1{\vbox{\offinterlineskip\hrule\halign{\vrule\enspace
$##$\hfil\strut\enspace\vrule&&\enspace$##$\hfil\enspace\vrule\cr#1}\hrule}}

\def\x{\hbar \omega \over kT}
\def\bx{\vbox{\hrule\hbox{\vrule\kern6pt\vbox{\kern6pt}\vrule}\hrule}}
\def\ctablel#1{\vbox{\offinterlineskip\hrule\halign{\vrule\enspace
$##$\hfil\strut\enspace\vrule&&\enspace$##$\hfil\enspace\vrule\cr#1}\hrule}}
~\vskip -3cm
$$\eqalignno{&&\ctablel{
{\rm Published~in~ International~Journal~of~Modern~Physics\,A~}{\bf 11}~{\rm 3667-3688~(1996)}\cr
}
\cr}$$
\bigskip 
\bigskip
\beginCenter
\Title
PAIRED ACCELERATED FRAMES\Footnote{$^\dagger$}{Based on a `raporteur report' 
of the
session ``Quantum Radiation and Accelerated Frames'' originally published in
the Proc. of the Seventh Marcel Grossmann Meeting on General Relativity,
Stanford, July 24-30, 1994, edited by R.T. Jantzen and M. Keiser (World 
Scientific, Singapore, 1996), 957-976.}
\endTitle
\beginAuthors
\Author{ULRICH H. GERLACH}
Department of Mathematics, Ohio State University, 
Columbus, OH 43210, USA
\endAuthor
\endAuthors
\endCenter

\Abstract{%
The geometrical and quantum mechanical basis for Davies' and Unruh's
acceleration temperature is traced to a type of quantum mechanical
(``achronal'') spin. Its existence and definition are based on pairs of
causally disjoint accelerated frames. For bosons the expected spin vector of
monochromatic particles is given by the ``Planckian power'' and the
``r.m.s. thermal fluctuation'' spectra. Under spacetime translation the spin
direction precesses around that ``Planckian'' vector. By exhibiting the
conserved achronal spin four-current, we extend the identification of achronal
spin from single quanta to multiparticle systems. Total achronal spin
conservation is also shown to hold, even in the presence of quadratic
interactions.
\hfil\break
\indent
In addition, the Lorentz invariance of the acceleration temperature is made 
explicit by the
introduction of pairs of ``spherical'' Rindler frames.
}%
\typeout{ >> MG7tex note: Footnotes in the abstract are bad form, no?}
\vskip .5truecm
Classical mechanics was the starting point by which Newton and
Einstein brought gravitation into our grasp. Newton's particle
trajectories carry the imprint of gravitational forces from which one
can deduce the properties of gravitation and which hence led to
Newton's equations for the gravitational potential. Einstein's
worldlines, via geodesic deviation, carry the imprint of spacetime
being curved, from which one can deduce gravitation as a manifestation
of geometry, and which hence inevitably leads to Einstein's equations
for the dynamics of the geometry.

The actual world is however quantum mechanical in nature, and the
starting point for gravitation should reflect this fact. Thus the
imprints of gravitation should not be carried by classical particle
worldlines but, more fundamentally, by wave functions or operators of
quantum mechanics instead.

\section{1. The Acceleration Temperature}

Nowadays, except for the energy $E =\hbar \omega$, it is difficult to
point to a purely quantum mechanical formula which is simpler
than that of Davies-Unruh (D-U) for the acceleration temperature [1,2]
$$\eqalignno{kT={\hbar\over c}{g\over 2\pi}.&&\hbox{(1.1)}\cr}$$
Can one therefore claim that in some way the simplicity of the D-U formula 
expresses a correspondingly fundamental aspect of nature? [``fundamental''
meaning that, given the context of our present day knowledge, this aspect 
serves to organize the largest possible set of physical
properties and phenomena into a readily identifiable existent]

The first thing one notices about the acceleration temperature is its ubiquity:
the quantum dynamics of all relativistic wave equations that have been examined
so far, be they for bosons [1,2,3], fermions [4,5], or for a nonlinear
field [6], give rise to this temperature.

The second thing to notice is that it also occurs in gravitation: equate
the acceleration $g$ to the surface gravity of a black hole event horizon,
$$\eqalignno{
g&={c^4 \over {4MG}}\quad . & (1.2) \cr }$$
With the gravitational constant $G$ introduced in this fashion, the 
acceleration temperature becomes the black hole temperature given by the
Bekenstein-Hawking formula.

The third thing to notice is that the occurrence of the two constants of nature
(the quantum of action $\hbar$, and the speed of light $c$) in the acceleration
temperature somehow expresses an amalgamation of quantum mechanics with the
causal (``light cone'') structure of spacetime. Indeed, that causal structure
is the geometrical corner stone of the D-U formula. It is expressed by the {\it
pair} of Rindler reference frames ($\xi >0$),
$$\eqalignno{\left.\matrix{t-t_0 =\pm\xi\sinh g\tau\cr  
z-z_0  = \pm\xi  \cosh g\tau\cr}
\right\}\quad \matrix {+:~~\hbox{``Rindler Sector I''}\cr -:
~~\hbox{``Rindler Sector II''}\cr}&&(1.3)}$$ 
so that
$$\eqalignno{ 
(z-z_0)^2 -(t-t_0)^2 &= \xi ^2 \quad .& (1.4)\cr }$$ 
They accelerate linearly into opposite directions relative to the reference 
event $(t_0,x_0)$, and they each give rise to the same static spacetime 
geometry, which is expressed by
$$\eqalignno{
ds^2&= -\xi^2 g^2 d\tau^2 + d\xi^2 + dy^2 +dx^2 \quad .
&(1.5)\cr}$$
This partitioning of spacetime into acceleration-induced pairs of adjacent
but causally disjoint (i.e. ``achronally'' related) Rindler frames is expressed
quantum mechanically by ``achronal spin''. This is done below.

\section{2. The Einstein Field Equations}

The nature of gravitation, i.e. the effect that it has on matter, as well as
the manner in which matter determines gravitation, is expressed by the Einstein
field equations. In their integral form they state that [7,8]
$$\eqalignno{
\left( 	\matrix {\hbox {``moment of} \cr
         \hbox { rotation''} \cr} \right)     
      = {8\pi G}~ \hbox {(``energy-momentum'')} &&\cr
}$$
More precisely, the right hand side is the amount of energy and momentum per 
unit three-volume. The left hand side, on the other hand, is the moment of 
rotation associated with the six faces of this three-volume.

These equations depend on matter being characterized in terms of energy and
momentum. To appreciate the significance of such a characterization, it is
necessary to recall an important fact about dynamical variables in general, and
energy and momentum in particular: They are of course attributes of single
particles, and they are conserved for free particles in the absence of
gravitation.  However, equally or even more important for our purpose is the
fact that {\it they indicate and imply the nature of the coordinate frame.}

If one chooses to characterize matter in terms of energy and momentum, (and
this is what Einstein did when he developed his theory of gravitation) then
one has implicitly committed himself to the use of instantaneous inertial
(``free float'') frames of reference at each event. In other words,
energy-momentum is the physical expression of the translation invariance of
free float frames. Thus the use of energy and momentum as dynamical variables
is predicated on a picture of spacetime consisting of a locally translation
invariant inertial frame (in geometry: ``tangent space'' with a basis)
attached to each event. Once such a commitment has been made, geometry forces
us to characterize these ``tangent spaces'' in terms of a metric, parallel
transport, curvature, and so on, as one proceeds towards arriving at the
Einstein's field equations.

To summarize, the Einstein field quations imply inertial frames of
reference. The spacetime arena for the physics of matter is to be viewed in
terms of ``free float frames'', one copy attached to each event.

\section{3. Inertial Frames vs Pairs of Accelerated frames}

The acceleration temperature implies a picture of spacetime which is in direct
conflict with the picture implied by the Einstein field equations. The former
is based on pairs of causally disjoint accelerated frames. The latter is based
on instantaneous inertial frames of reference. Thus one has the peculiar state
of affairs that the acceleration temperature implies pairs of accelerated 
frames, while the field equations imply inertial frames.

Even though both types of frames have a Cauchy hypersurface for initial value
data of the relativistic equations of for matter, the difference in the
classical as well as the quantum dynamics is irreconcilable. A pair of
accelerated frames cannot be deformed into an inertial frame and vice
versa. Ultimately one has to make a choice: pairs of accelerated frames implied
by quantum mechanics, or inertial frames implied by the Einstein field
equations.

\section{4. Lorentz Invariance}
If the acceleration temperature is to play a fundamental role in
physics, then it must be in harmony
with Lotentz invariance. This clearly does not seem to be the case: there is a
preferred spatial direction along which the linear acceleration takes place.
Moreover, the spacetime framework expressed by Eqs.(1.3)-(1.5) expresses
another related deficiency: The pair of accelerated frames single out a 
reference event
$(z-z_0=t-t_0=0)$ in the longitudinal spacetime plane, but the location of the
frames in the transverse ($x\hbox {-y}$) plane is left totally ambiguous.

However these
deficiencies are removed by a frame which is characterized by accelerating
spherical coordinates
$$\eqalignno{
t-t_0 &= \xi \sinh g\tau \cr
z-z_0 &= \xi \cosh g\tau \cos \theta \cr
y-y_0 &= \xi \cosh g\tau \sin \theta \sin \phi \cr
x-x_0 &= \xi \cosh g\tau \sin \theta \cos \phi\quad .&(4.1)\cr}$$
The Minkowski metric relative to these coordinates is
$$\eqalignno{
ds^2&= -\xi^2 g^2 d\tau^2 + d\xi^2 + \xi^2 \cosh ^2 \tau
(d\theta^2 + \sin^2 \theta d \phi^2).
&\cr}$$
They coordinatize the Lorentz invariant histories of (fictitious)
``bubbles'' contracting and reexpanding around their reference event $(t_0,z_0,y_0,x_0)$. See Figure 1. 

\epsffile[-55 0 300 300]{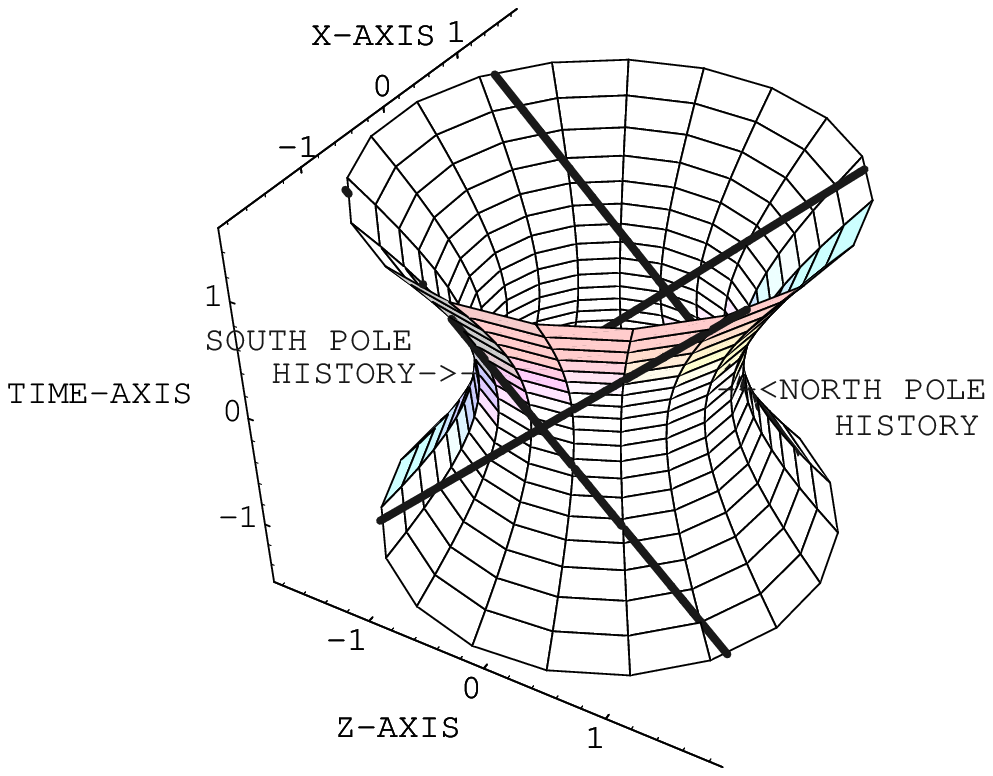}
\medskip
\cli{\bf FIGURE 1}
\medskip
{\narrower {\narrower\smallskip\noindent
Lorentz invariant history of a contracting and reexpanding ``bubble''
partitioned by future and past event horizons (the heavy forty five degree
lines) into a pair of causally disjoint (=``achronally'' related) Rindler
Sectors I and II. The y-dimension has been suppressed.  The south and the north
pole of the bubble trace out hyperbolic world lines in the respective Rindler
Sectors I and II on the x=0 plane. The history of the ergosphere consists of
those events on the hyperboloid which lie between the event horizons and the
two respective hyperbolas $(z-z_0)^2-(t-t_0)^2=\xi ^2 \cos ^2 \theta ,~ 
\pm \vert x-x_0\vert =
\xi \sin \theta $, with vertices determined by Eq.(4.10).
\smallskip}}
\medskip

The
coordinates of this frame imply that each bubble has a hyperbolic history,
$$\eqalignno{ 
(x-x_0)^2 +(y-y_0)^2 +(z-z_0)^2 -(t-t_0)^2 &= \xi ^2 \quad ,&(4.2) \cr }$$ 
which is invariant under the actions of the Lorentz group
SO(3,1) around the event $(t_0,z_0,y_0,x_0)$. This is a considerable
improvement over the hyperbolic history 
$$\eqalignno{ 
(z-z_0)^2 -(t-t_0)^2 &= \xi ^2 & \cr }$$ 
of a linearly accelerated frame, which lacks geometric isotropy, and which is 
(``boost'') invariant under the actions of $SO(1,1)$ only. Furthermore, there
is no ambiguity as to the spacetime location $(t_0,z_0,y_0,x_0)$ of the
spherically symmetric frame.

However the key feature of this frame is quantum mechanical: each bubble is
heated [9] to the acceleration temperature given by the D-U formula
$$\eqalignno {
kT= {\hbar \over c} {g \over {2\pi}}\quad ;&& (4.3)\cr }$$
that is to say, a thermometer comoving with this bubble registers the 
acceleration
temperature if the relativistic wavefield is in the Minkowski vacuum
state.

Two geometric properties are responsible for this temperature: (a) the
existence of a frame which is time invariant so that a relativistic wave system
can come to thermal equilibrium at its acceleration temperature, and (b) the
existence of past and future event horizons which partition the quantized
relativistic wave field into a pair of causally disjoint subsystems.

\bigskip
\n {\it 4.1~~~Spherical Rindler Frame}
\medskip

With any Lorentz invariant ``contracting and expanding bubble'' frame
coordinatized by Eq.(4.1) there is a concomitant unique frame with the required
properties (a) and (b). This frame is coordinatized by the {\it {spherical
Rindler coordinates}} [10],

$$\eqalignno{
\left.\matrix  {
t - t_0 =& \xi \cos \theta \sinh \tau  \cr
z - z_0 =& \xi \cos \theta \cosh \tau \cr
y - y_0 =& \xi \sin \theta \sin \phi \cr
x - x_0 =& \xi \sin \theta \cos \phi \cr}\right \}
\left \{\matrix { {\rm Rindler~ Sector~ I} &: 0 \le \theta < {\pi \over 2} \cr
{\rm Rindler~ Sector~II} &: {\pi \over 2} < \theta \le \pi , \cr}\right. 
&& (4.4)\cr }$$
The past and future event horizons are 
$$\eqalignno{
z-z_0 &=\pm |t-t_0|\cr
y-y_0&=\xi \sin \phi\cr
x-x_0&=\xi \cos \phi , \quad  \quad 0<\xi <\infty , 0\le\phi <2\pi \quad .
&(4.5)\cr}$$
They partition the history of each concentric bubble into what for a linear
Rindler frame corresponds to the pair of Rindler Sectors I and II in Eq.(1.3).
See Figure 1.
The north pole ($\theta =0$) of each sphere has a hyperbolic world line in
Rindler Sector I, while the south pole ($\theta = {\pi }$) has a
corresponding world line in Rindler Sector II. The metric on each of
these Rindler sectors is
$$\eqalignno{
ds^2&= -\xi^2 \cos^2 \theta d\tau^2 + d\xi^2 + \xi^2 
(d\theta^2 + \sin^2 \theta d \phi^2),
&{(4.6)}\cr}$$
As required, the spherical Rindler frame is independent of time. The 
singularity in the metric
$$\eqalignno{
\xi ^2 \cos ^2 \theta = 0 \quad , && \cr } $$
expresses the set of events where  the past and the future event horizons 
intersect.
These events are on the equatorial plane $z-z_0=0~(\theta ={\pi \over 2})$ 
of each contracting and reexpanding bubble at $t-t_0=0$, the moment of time 
symmetry as seen in the globally inertial frame $(t,z,y,x)$.
These {\it bifurcate equatorial events} are the starting (or termination) 
events of the null ray histories of
two swarms of null particles which travel into opposite directions
parallel to the $z$-axis. The collection of these null ray histories make up
the {\it future} and the {\it past event horizons}, Eq.(4.5).
They form the boundaries of four mutually exclusive and jointly exhaustive
coordinate neighborhoods on each hyperbolic ``bubble'' history, Eq.(4.2).
They are the four Rindler Sectors (``wedges''): I and II,
given by Eqs.(4.4), and the chronological future (F) and past (P), given by 
$$\eqalignno{
\left.\matrix  {t -t_0 =& \xi \sinh \theta \cosh \tau  \cr
z - z_0 =& \xi \sinh \theta \sinh \tau \cr
y -y_0 =& \xi \cosh \theta \sin \phi \cr
 x-x_0 =& \xi \cosh \theta \cos \phi \cr}\right \}
\left \{\matrix { {\rm Rindler~ Sector~ F} &: 0< \theta < \infty \cr
{\rm Rindler~ Sector~P} &: -\infty < \theta <0  . \cr}\right. 
&& (4.7)\cr }$$
The metric on each of these two sectors is
$$\eqalignno{
ds^2&= \xi^2 \sinh^2 \theta d\tau^2 + d\xi^2 + \xi^2 
(-d\theta^2 + \cosh^2 \theta d \phi^2),
&{(4.8)}\cr}$$
\bigskip
\n {\it 4.2~~~Rotating Spherical Rindler Frame and its Ergosphere}
\medskip

The spherical Rindler coordinates associated with a set of contracting and
reexpanding bubbles is not unique. It depends on the choice of a common axis of
rotation around the chosen north pole and south pole of each bubble. This
choice is equivalent to the choice of an equator. It determines the location of
the north and south poles and hence the respective sectors I and II of the
spherical Rindler frame. Put differently, there is one-to-one correspondence
between a given set of bifurcate equatorial events and a spherical Rindler
frame.

The existence of an axis of rotation permits the consideration of a spherical
Rindler frame which rotates. This rotation is superimposed on the accelerative
contraction and expansion of each bubble. If the common angular velocity is
$\Omega$, then upon introducing the corotating angular coordinate
$$\eqalignno{
\tilde \phi =\phi-\Omega \tau, && \cr
}$$
the metric, Eq.(4.6), becomes
$$\eqalignno{
ds^2&= -\xi^2 (\cos^2 \theta -\Omega ^2 \sin ^2 \theta) d\tau^2 + 
2 \Omega \xi ^2 \sin ^2 \theta d\tilde \phi d\tau +
d\xi^2 + \xi^2 (d\theta^2 + \sin^2 \theta d \tilde \phi^2).
& \cr
&&{(4.9)}\cr}$$
The ergosphere [11] is that region of space where the otherwise timelike 
Killing vector 
field $\partial \over {\partial \tau}$ is spacelike. For a spherical 
Rindler frame with rotation the ergosphere is an equatorial belt. Its
latitudinal thickness is given by
$$\eqalignno{
-\Omega <{{\cos \theta} \over {\sin \theta}}<\Omega && (4.10) \cr
}$$
independent of the radius of each sphere.

In spite of their geometrical difference, the $SO(1,1)$-based linear Rindler
coordinate frames, Eq.(1.3), and the $SO(3,1)$-based spherical Rindler
coordinate frames, Eq.(4.4), have a fundamental property in common: they both
organize events into pairs of achronally related (``causally disjoint'')
spacetime regions I and II.

Such a partitioning into pairs introduces a qualitatively new feature into the
quantum mechanics of a single particle, many particles (``quantum field
theory''), as well as processes of interaction between them. This feature is
achronal spin, which is developed below. The qualitative idea of achronal spin
is easy to state. However its exact statement is in terms of the normal mode
technology relative to Rindler sectors I and II. That technology is
as-yet-undeveloped for spherical Rindler frames. Thus we shall use as a proxy
the well-developed technology for pairs of linearly accelerated Rindler frames
instead. This means that Lorentz invariance gets sacrificed in favor of
simplicity. However the essential feature, achronal spin, is still present.
\bigskip

\section{5. Achronal Spin}

Consider the picture of spacetime which consists of pairs of oppositely
accelerated reference frames, one pair attached to each event.

\bigskip
\n {\it 5.1~~~Existence}
\medskip

The physical consequences of such a geometrical picture is that the complete
set of quantum mechanical observables of a particle governed by a linear
relativistic wave equation has a qualitatively new member, ``achronal spin''.

Achronal spin is a quantum mechanical expression of acceleration-induced
partitioning of spacetime into pairs of adjacent but causally disjoint (i.e.
``achronally'' related) Rindler sectors.
Achronal spin is a conserved observable with a discrete spectrum, but,
being a member of a complete set of observables, also indicates the pair of
causally disconnected Rindler sectors at each reference event.

Assume for the sake of simplicity that these Rindler sectors result from a pair
of frames accelerating linearly. Thus the two transverse dimensions become
trivial and the spacetime arena becomes the two dimensional Lorentz plane.  The
boost invariance of the wave equation on this plane implies that the familiar
boost energy is a conserved observable. 

However there clearly has to be another observable: The relative phase and
amplitude of the wave function in the two Rindler Sectors I and II can be
changed at will without changing the boost energy and without destroying the
fact that the wave function is a solution. In other words, the causally
disjoint nature of the two Rindler domains I and II implies that a solution to
the relativistic wave equation can be transformed into another solution by
simply changing the relative phase and amplitude of the wave function defined
on the two domains.

These changes form therefore an invariance group of the relativistic wave
equation. This invariance group obviously commutes with Lorentz boosts. It
follows that the generators of this group are observables which are also
conserved. The two most important questions are: (1) What are the values of
these observables? and (2) What is the ``complete set of commuting
observables'' for the wave equation relative to the pair of Rindler frames I
and II?

The answers to both questions follow from the representation and the Lie
algebra of the invariance group. Consider a single scalar charge. The 
wave function on which the group
acts has two components because the solution is defined on the two disconnected
Rindler sectors, Eq.(1.3). Let this two component wave function be 
$$\eqalignno{\psi (\tau ,\xi )=\left[\matrix {\psi_{\hbox{I}}(\tau ,\xi )\cr 
\psi_{\hbox{II}}(\tau ,\xi)\cr} 
\right] \quad ,&&(5.1)\cr}$$
It satisfies the Klein Gordon equation in I and in II:
$$\eqalignno{\left[-{1\over {\xi ^2}}{\partial^2\over \partial\tau^2}+{1\over
\xi}{\partial\over\partial\xi}\xi{\partial\over \partial\xi}
-k^2\right]
\left[\matrix {\psi_{\hbox{I}}(\tau ,\xi )\cr 
\psi_{\hbox{II}}(\tau ,\xi)\cr}\right]
=0 \quad .&&(5.2)\cr}$$ 
Here $k^2=k^2_x+k^2_y+{m^2c^2\over\hbar^2}$. The basic idea is illustrated
by considering  plane wave states
$$\eqalignno{
e^{\mp i(t-t_0)k\cosh \theta +i(x-x_0)k\sinh \theta} &&(5.3) \cr
}$$
of either positive (upper sign) or negative (lower sign) Minkowski frequency.
These states, we recall, are parametrized by the ``pseudo'' angular parameter 
$\theta$ on the mass
hyperboloid $\omega ^2_k-k^2_z=k^2_x+k^2_y + {m^2 \over \hbar ^2} \equiv k^2$,
$$ \eqalignno {
\omega _k&=k \cosh \theta >0\cr
     k_z &=k \sinh \theta & (5.4) \cr
}$$
In terms of the boost invariant normal modes 
$K_{i\omega}(k\xi )e^{-i\omega\tau}$,
a plane wave state has an expansion consisting of two components, one for each
of the two Rindler domains $I$ and $II$, Eq.(1.3), on which it is defined. 
The result is [12]
$$\eqalignno{
\int\limits^{\infty}_{-\infty}
\left[\matrix {e^{\pi\omega/2}\cr e^{-\pi\omega/2}\cr}\right]
{K_{i\omega}(k\xi )\over \pi}
e^{-i\omega\tau}e^{i\omega\theta} d\omega
&=\left[\matrix {e^{-i(t-t_0)k\cosh \theta +i(x-x_0)k\sinh \theta}\vert _I\cr
                 e^{-i(t-t_0)k\cosh \theta +i(x-x_0)k\sinh \theta}\vert _{II}
                                                                \cr}\right] 
&(5.5a)\cr
&=\left( \matrix {\hbox{positive frequency}\cr
                  \hbox{plane wave state}\cr}  \right)
\equiv \int\limits ^\infty _{-\infty} \psi ^+_\omega d\omega \cr
\noalign{\hbox{and}}
\int\limits^{\infty}_{-\infty}
\left[\matrix {e^{-\pi\omega /2}\cr 
               e^{\pi\omega/2}\cr}\right]
{K_{i\omega}(k\xi )\over \pi} 
e^{-i\omega\tau}e^{i\omega\theta} d\omega
&=\left[\matrix {e^{i(t-t_0)k\cosh \theta -i(x-x_0)k\sinh \theta}\vert _I\cr
                 e^{i(t-t_0)k\cosh \theta -i(x-x_0)k\sinh \theta}\vert _{II}
                                                                \cr}\right] 
&(5.5b)\cr
&=\left( \matrix {\hbox{negative frequency}\cr
                  \hbox{plane wave state}\cr} \right)
\equiv \int\limits ^\infty _{-\infty} \psi^- _\omega d\omega \cr
}$$

Quantum mechanics demands that the action of the invariance group be such
that the (Klein Gordon) inner product, Eq.(5.9) below, between quantum states 
remains invariant.
Applied to each spectral component, this invariance determines the three one
parameter subgroups
$$\eqalignno{
\left[ \matrix {\cosh {\alpha \over 2} &i\sinh {\alpha \over 2}\cr
-i\sinh {\alpha \over 2} &\cosh {\alpha \over 2}\cr}\right]
&\equiv e^{iL_1 \alpha} \quad ,&(5.6a)\cr
\left[ \matrix {\cosh {\theta \over 2} &\sinh {\theta \over 2}\cr
\sinh {\theta \over 2} &\cosh {\theta \over 2}\cr}\right]
&\equiv e^{iL_2 \theta} \quad ,&(5.6b)\cr
\left[ \matrix {e^{\phi \over 2} &0\cr 0&e^{\phi \over 2}\cr}\right]
&\equiv e^{iL_3 \phi} \quad .&(5.6c)\cr
}$$
Here
$$\eqalignno{
\ora L:\{L_1,L_2,L_3\}=\left\{{1\over 2}\left[\matrix {0&1\cr
-1&0\cr}\right],{1\over 2}\left[\matrix {0&-i\cr -i&0\cr}\right],{1\over
2}\left[\matrix {1&0\cr 0&-1\cr}\right]\right\} &&(5.7)\cr}$$
are one half the modified Pauli spin matrices. They are the three {\it
achronal spin} operators. They are conserved, and they
generate the invariance group, which evidently is $SU(1,1)$. They 
satisfy the commutation relations
$$\eqalignno{
[L_1,L_2]&=-iL_3\cr
[L_2,L_3]&=iL_1\cr
[L_3,L_1]&=iL_2\quad ,&(5.8)\cr}$$
as well as
$$\eqalignno{
\ora L\cdot\ora L&\equiv L^2_3-L^2_2-L^2_1\left(={1\over 2}\left({1\over
2}+1\right)\right).&\cr
}$$

One concludes that (1) {\it a scalar charge has achronal spin $1\over 2$} and
that (2) relative to a pair of Rindler frames in the two dimensional Lorentz
plane the complete set of commuting observables for such a charge consists of
the boost energy {\it and} the (third component of) achronal spin, $\{ 
i{\partial
\over {\partial \tau} }, L_3\}$.
\bigskip
\n {\it 5.2~~~Matrix Elements}
\medskip

For any observable, including achronal spin, matrix elements are of particular
physical interest: Among other things, they yield, of course, the expected spin
orientation for a given quantum state. 

This leads to a striking encounter between group theory and thermodynamics:
The achronal spin expectation value determines a direction whose
vertical component is ``zero point energy'' plus ``Planckian power''
while its horizontal component is the ``r.m.s. thermal fluctuation''. Nature is
trying to tell us that quantum mechanical symmetries of spacetime manifest
themselves thermodynamically through Boltzmann's maximum probability
principle.  

Matrix elements are based on a quantum mechanical inner product, which for a
relativistic scalar charge is that of Klein Gordon [13],
$$\eqalignno{
\langle \psi,\psi'\rangle&={i\over 2}\int^0_{\infty}
\left(\psi^{\ast}_{II}{\partial\over
\partial\tau}\psi'_{II}-{\partial\over\partial\tau}\psi^{\ast}_{II}
\psi'_{II}\right){d\xi\over\xi}+{i\over
2}\int^{\infty}_0\left(\psi^{\ast}_{I}{\partial\over\partial\tau}
\psi_I'-{\partial\over\partial\tau}\psi^{\ast}_I\psi_I\right){d\xi\over\xi}
\cr
&={i\over
2}\int^{\infty}_0\left(\psi^{\dag}\sigma_3{\partial\over\partial\tau}
\psi'-{\partial\over\partial\tau}\psi^{\dag}\sigma_3\psi'\right)
{d\xi\over\xi}.&(5.9)\cr}$$
Here $\sigma _3$ is the Pauli matrix
$$\sigma_3=\left[\matrix {1&0\cr 0&-1\cr}\right] \quad .$$
Let us determine all the achronal spin matrix elements 
$\langle \psi _\omega ^+,\ora L\psi^+ _{\omega '}\rangle$,
$\langle \psi _\omega ^+,\ora L\psi^- _{\omega '}\rangle$, etc.
for the pairs of
positive and negative Minkowski frequency boost eigenstates $\psi^+ _\omega$ 
and$\psi^- _\omega$ in Eq.(5.5).
Introducing the ``zero point energy'' plus ``Planckian power'' spectrum
$$\eqalignno{
{{\cosh \pi \omega }\over {2 \sinh \pi \omega}} &={1\over 2}+{1\over {\exp 2 \pi \omega -1} }\cr
                            &\equiv{1\over 2}+N &\cr
\noalign{\hbox{and the ``r.m.s. thermal fluctuation'' spectrum}}
{1\over {2 \sinh \pi \omega}}&=\sqrt{N(N+1)}\cr
&\cr}$$
one obtains
$$\eqalignno{
\left[ \matrix {\langle \psi _\omega ^+,L_1 \psi^+ _{\omega '}\rangle 
               &\langle \psi _\omega ^+,L_1 \psi^- _{\omega '}\rangle\cr 
                \langle \psi _\omega ^-,L_1 \psi^+_{\omega '}\rangle
               &\langle \psi _\omega ^-,L_1 \psi^- _{\omega '}\rangle\cr}
                                          \right] 
&=
~~~\left[ \matrix {\sqrt{N(N+1)}
               &{1\over 2}+N\cr 
                {1\over 2}+N
               &\sqrt{N(N+1)}\cr}  \right] \delta (\omega-\omega ') &\cr
\left[ \matrix {\langle \psi _\omega ^+,L_2 \psi^+ _{\omega '}\rangle 
               &\langle \psi _\omega ^+,L_2 \psi^- _{\omega '}\rangle\cr 
                \langle \psi _\omega ^-,L_2 \psi^+_{\omega '}\rangle
               &\langle \psi _\omega ^-,L_2 \psi^- _{\omega '}\rangle\cr}
                                          \right] 
&=~~~~~~~~~~~
\left[ \matrix {0
               &{i\over 2}\cr 
                {-i\over 2}
               &0   \cr}  \right] \delta (\omega-\omega ') &(5.10)\cr
\noalign {\hbox {and}}
	\left[ \matrix {\langle \psi _\omega ^+,L_3 \psi^+ _{\omega '}\rangle 
               &\langle \psi _\omega ^+,L_3 \psi^- _{\omega '}\rangle\cr 
                \langle \psi _\omega ^-,L_3 \psi^+_{\omega '}\rangle
               &\langle \psi _\omega ^-,L_3 \psi^- _{\omega '}\rangle\cr}
                                          \right] 
&=
~~~\left[ \matrix {{1\over 2}+N
                &\sqrt{N(N+1)}\cr
                \sqrt{N(N+1)}
                &{1\over 2}+N\cr}  \right] \delta (\omega-\omega ') &\cr
\noalign {\hbox {while the inner products are simply}}
\left[ \matrix {\langle \psi _\omega ^+, \psi^+ _{\omega '}\rangle 
               &\langle \psi _\omega ^+, \psi^- _{\omega '}\rangle\cr 
                \langle \psi _\omega ^-, \psi^+_{\omega '}\rangle
               &\langle \psi _\omega ^-, \psi^- _{\omega '}\rangle\cr}
                                          \right] 
&=~~~~~~~~~~~
\left[ \matrix {1&0\cr 
                0&-1\cr}  \right] \delta (\omega-\omega ') &(5.11)\cr
}$$

\subsection{5.3 Planckian Spectral Vector}
Relative to the quantum states (``spinors'') $\psi ^+
_\omega$, as well as relative to $\psi ^- _\omega$, the expectation value of
the achronal spin operator {\bf L} is
$$\eqalignno {
\langle\psi^\pm_\omega,{\bf L}\psi^\pm_{\omega'} \rangle
&\equiv {1\over 2} {\bf l}^\pm _\omega ~~\delta (\omega-\omega ') &(5.12)\cr
&=\lbrace \left[{1\over\exp {\x}-1}+{1\over (\exp {\x}
-1)^2}\right]^{{1\over 2}},0,{1\over 2}+{1\over \exp {\x} -1}\rbrace 
~~\delta (\omega-\omega ')
\quad .\cr
}$$
These are the diagonal matrix elements. The vector ${\bf l^+_\omega}$, as well 
as ${\bf l^-_\omega}$, formed from these expectation values is a real
three-dimensional vector, which we shall call the ``Planckian spectral 
vector''. This vector is unique. It is the same for all plane wave states,
and its components are two of Nature's most ubiquitous spectra: the
``zero point energy'' plus ``Planckian thermal power'' spectrum for the third
component and the ``r.m.s. thermal fluctuation'' spectrum for the first 
component. The Planckian spectral vector has a dynamical significance which
is described in the next subsection.

The off-diagonal elements are complex. Their real and imaginary parts
$$\eqalignno {
\langle\psi^\mp_\omega,{\bf L}\psi^\pm_{\omega'} \rangle
&=Re~\langle\psi^\mp_\omega,{\bf L}\psi^\pm_{\omega'} \rangle
+i~Im~  \langle\psi^\mp_\omega,{\bf L}\psi^\pm_{\omega'} \rangle \cr
&\equiv ({1\over 2} {\bf m}^\pm _\omega + i {1\over 2} {\bf n}^\pm _\omega )
~~\delta (\omega-\omega ') \cr
}$$
yield two addtional real three-dimensional vectors,
$$\eqalignno {
{\bf m}^\pm _\omega &=
2\lbrace {1\over 2}+{1\over \exp {\x} -1},0,\left[{1\over\exp {\x}-1}+ 
                    {1\over (\exp {\x}-1)^2}\right]^{{1\over 2}}\rbrace , \cr
{\bf n}^\pm _\omega &=
\pm \{ 0,1,0\} \quad . &(5.13)\cr
}$$
Together the vectors ${\bf l}^+_\omega$, ${\bf m}^+ _\omega$,
and ${\bf n}^+ _\omega$ form a triad, which is ``Lorentz'' orthonormal in the
sense that 
$$\eqalignno{
{\bf l}^+ _\omega \cdot {\bf l}^+ _\omega 
&\equiv (l_{\omega 3})^2- (l_{\omega 2})^2-(l_{\omega 1})^2  \cr
&=1 \cr
{\bf m}^+ _\omega \cdot {\bf m}^+ _\omega 
&={\bf n}^+ _\omega \cdot {\bf n}^+ _\omega =-1 \cr
{\bf l}^+ _\omega \cdot {\bf m}^+ _\omega & =
{\bf l}^+ _\omega \cdot {\bf n}^+ _\omega =
{\bf l}^+ _\omega \cdot {\bf n}^+ _\omega =0 \quad . & (5.14) \cr
}$$
The directions of the vectors making up this triad are unique. Like the
Planckian spectral vector, they are the 
same for all plane wave states and their components are fixed by Nature. 
\bigskip
\subsection{5.4 Precession under Spacetime Translation}

We now ask and answer two questions: (1) How does achronal spin change in the
paired Rindler frame attached to a given fixed event? (2) How does achronal
spin change under spacetime tranlations from the paired frame at one event to
the paired frame at another?

The quick answer to the first question is that it does not change at all; it 
remains fixed under boost time evolution. The answer to the second question is 
that it precesses around the direction of the Planckian spectral vector.

Let us see how the Klein-Gordon wave equation furnishes the answer to these
two questions. Consider the expectation value taken with respect to an 
arbitrary Klein-Gordon state,
$$\eqalignno{
\psi &=
       \int\limits^{\infty}_{-\infty}
                      \left[\matrix { a_\omega \cr b^*_\omega \cr}\right] 
{K_{i\omega}(k\xi )\over \pi} e^{-i\omega\tau} d\omega & (5.15) \cr
&\equiv \int\limits^{\infty}_{-\infty}  \psi_\omega d\omega \quad .\cr
}$$
More precisely, take the expectation values of ${\bf L}:(L_1,L_2,L_3)$
with respect to the spectral components $\psi _\omega$. Being based on the 
standard quantum mechanical (Klein-Gordon) inner product, Eq.(5.9)
$$\eqalignno{
\langle \psi _\omega ,\psi _{\omega '} \rangle &=
(a^{\ast}_{\omega} a_{\omega} -b_{\omega} b^{\ast}_{\omega})
{{\delta (\omega-\omega')}\over {2\sinh \pi \omega}}  
\quad ,&(5.16)  \cr
\noalign {\hbox {these expectation values are (by inspection of Eq.(5.7))}}
\cr
\langle \psi _\omega ,L_1 \psi _{\omega '}\rangle &=
{1\over 2}(a_{\omega} b_{\omega} +a^*_{\omega} b^*_{\omega}) 
      {{\delta (\omega-\omega')}\over {2\sinh \pi \omega}}  \cr
\langle \psi _\omega ,L_2 \psi _{\omega '}\rangle &=
-{i\over 2}(a_{\omega} b_{\omega} -a^*_{\omega} b^*_{\omega}) 
      {{\delta (\omega-\omega')}\over {2\sinh \pi \omega}}  \cr
\langle \psi _\omega ,L_3 \psi _{\omega '}\rangle &=
~~{1\over 2}(a^{\ast}_{\omega} a_{\omega} +b_{\omega} b^{\ast}_{\omega})
      {{\delta (\omega-\omega')}\over {2\sinh \pi \omega}}  
\quad .&(5.17)\cr
}$$

The time (i.e $\tau$, the ``boost'' time) independence of these expressions
imply the first result: {\it In any given paired accelerated frame achronal
spin ${\bf L}:(L_1,L_2,L_3)$ is a (set of) constant(s) of motion}. Although
not simultaneously measurable, they are nevertheless fixed and conserved. 
This answers the first question posed at the beginning of this section.

What about the second question, change of achronal spin under spacetime
translation? There is a definite answer, ``precession''. It is illustrated by
considering what in a global inertial frame is an arbitrary superposition of
(i) a positive Minkowski frequency (``particle'', (5.5a)) plane wave state
and (ii) another plane wave state of negative Minkowski frequency 
(``antiparticle'', (5.5b)).
Let the amplitudes of these states be $\tilde a$ and $\tilde b^*$ 
respectively [15]:
$$\eqalignno{
\psi&=
\tilde a \int\limits^{\infty}_{-\infty}
e^{i\omega \theta _1} 
\left[\matrix {e^{\pi\omega/2}\cr e^{-\pi\omega/2}\cr}\right] 
{K_{i\omega}(k\xi )\over \pi} e^{-i\omega\tau} d\omega     +
\tilde b^* \int\limits^{\infty}_{-\infty}
e^{i\omega \theta _2} 
 \left[\matrix {e^{-\pi\omega/2}\cr e^{\pi\omega/2}\cr}\right]
{K_{i\omega}(k\xi )\over \pi}
e^{-i\omega\tau} d\omega \cr
&\equiv
\int\limits^{\infty}_{-\infty}
\lbrace 
\tilde a  
\psi ^+_\omega +
\tilde b^*  
\psi ^-_\omega
\rbrace d\omega  &(5.18) \cr
&\equiv \int\limits^{\infty}_{-\infty} \tilde \psi _\omega d\omega \cr
}$$
What is the spectral achronal spin direction when the charge is in this state,
and how does this direction change under spacetime translations? 
Before such a translation the expected spin direction is
$$\eqalignno {
       {{\langle\tilde\psi_\omega,{\bf L}  \tilde\psi _{\omega '}
\rangle}\over {\langle\tilde\psi_\omega,\tilde\psi_{\omega '}\rangle}}
=&{{\vert \tilde a \vert ^2 +\vert \tilde b \vert ^2}\over
   {\vert \tilde a \vert ^2 -\vert \tilde b \vert ^2} }
{{\bf l}_\omega\over 2}\cr
&+{{2\vert \tilde a \vert \vert \tilde b \vert }\over
   {\vert \tilde a \vert ^2 -\vert \tilde b \vert ^2} }
\left[
{{\bf m}_\omega\over 2}\cos \delta +
{{\bf n}_\omega\over 2} \sin \delta
\right].
&(5.19) \cr
}$$
Here $\delta =\delta_1 +\delta_2$ is the sum of the phase angles of the two 
amplitudes
$$\eqalignno {
\tilde a &= \vert \tilde a \vert e^{i\delta_1} \cr
\tilde b &= \vert \tilde b \vert e^{i\delta_2} & (5.20)\cr
}$$
Now subject the plane wave state in Eqs.(5.5) to a translation. This action
alters the basis spinors $[e^{\pi\omega/2}~~~e^{-\pi\omega/2}]^{transpose}$ in
Eq.(5.5a) by a common $\omega$-independent phase factor. When introduced into
Eq.(5.18) (in order to subject $\psi$ to the same translation) this phase
factor appears as a multiplier of the coefficient $\tilde a$ under the
integral. In other words, a translation is represented by multiplying the
amplitude $\tilde a$ by a common phase factor. Similarly, $\tilde b^*$ 
also gets multiplied by a (different) phase factor:
$$\eqalignno {
\tilde a &\rightarrow \tilde a e^{i\gamma_1} \cr
\tilde b &\rightarrow \tilde b e^{i\gamma_2} & (5.21)\cr
}$$
The effect of these spacetime translation-induced phase changes is to
rotate the expected achronal spin around the Planckian spectral direction
$$\eqalignno {
{\bf l} _\omega 
&=\lbrace \left[{1\over\exp {\x}-1}+{1\over (\exp {\x}
-1)^2}\right]^{{1\over 2}},0,{1\over 2}+{1\over \exp {\x} -1}\rbrace 
\quad &(5.22)\cr
}$$
by the angle $\gamma =\gamma_1 +\gamma_2$:
$$\eqalignno {
       {{\langle\tilde\psi_\omega,{\bf L}  \tilde\psi _{\omega '}
\rangle}\over {\langle\tilde\psi_\omega,\tilde\psi_{\omega '}\rangle}}
\rightarrow &{{\vert \tilde a \vert ^2 +\vert \tilde b \vert ^2}\over
   {\vert \tilde a \vert ^2 -\vert \tilde b \vert ^2} }
{{\bf l}_\omega\over 2}\cr
&+{{2\vert \tilde a \vert \vert \tilde b \vert }\over
   {\vert \tilde a \vert ^2 -\vert \tilde b \vert ^2} }
\left[
{{\bf m}_\omega\over 2}\cos (\delta +\gamma)+
{{\bf n}_\omega\over 2} \sin (\delta +\gamma)
\right].
&(5.23) \cr
}$$
The hyperbolic angle of 
inclination of the spin axis from the Planckian vector is given by
$$\eqalignno{
\tanh \alpha = 
{{2\vert \tilde a \vert \vert \tilde b \vert }\over
   {\vert \tilde a \vert ^2 +\vert \tilde b \vert ^2} }&&(5.24)\cr 
}$$
The angular amount by which the spin axis precesses around the Planckian
vector is proportional to the spacetime displacement 
$(\triangle t,\triangle x)$. Indeed, the plane wave states Eq.(5.5) imply 
that this precession angle is
$$\eqalignno{
\gamma \equiv \gamma_1 +\gamma_2
=\triangle t (\cosh \theta_1 +\cosh \theta_2)k-
\triangle x (\sinh \theta_1 +\sinh \theta_2)k &&(5.25)\cr
}$$
\vfill \eject

\section{6. Paired Accelerated Frames: Interacting Multiparticle Systems}

Gravitation manifests itself by the imprints it leaves on classical or, more
fundamentally, on quantum mechanical processes. The imprints on the orbits and
world lines of particles and even on wave fields are well-known and
appreciated. Such a harmonious and universally accepted assessment does not,
however, prevail for the imprints that gravitation leaves on individual
quanta. The problem is that, in the absence of appropriate symmetries,
gravitation introduces an ambiguity as to the distinction between positive and
negative frequency field oscillators. As a consequence, there is an ambiguity
as to the very definition of what is meant by a particle, i.e. a carrier of
energy and momentum. To tackle this conundrum, the viewpoint has been
advocated that a particle be identified by the fact that it brings about a
change in the state of a detecting apparatus which follows some prespecified
world line.

In practical terms this means that the onus of the ambiguity of the
negative-positive frequency distinction has been shifted to the arbitrariness
of the detector world line.

This state of affairs is somewhat reminiscent of the problem Euclidean
geometers faced before Riemannian geometry: ``Given a straight line
(``geodesic''), through a close-by point, draw another straight line parallel
line parallel to the first.'' In the absence of appropriate symmetries of the
manifold, a pair of straight lines can not be identified unambiguously as
having positive or negative relative slope. 
To deal with this problem Riemann's qualitatively different viewpoint
had to be taken.

\subsection {6.1 Binary Frame Picture of Spacetime}

One is led therefore to consider that the quantum mechanical
positive-negative frequency ambiguity introduced by gravitation requires an
analogous change in viewpoint, i.e. in the way we picture spacetime. In other
words, this ambiguity is evidence of a to-be-identified geometrical structure
which is a quantum mechanical carrier of the imprints of gravitation.

What is the framework which would permit the accomodation of such imprints?
The ubiquity of the acceleration temperature, Eq.(1.1), demands that this
framework consist of viewing spacetime in a qualitatively different way:
Instead of assigning to each event a locally inertial (``free float'')
reference frame (i.e. a tangent space with a basis), spacetime is to be 
pictured as a union
of pairs of accelerated reference frames, which are achronally related and
whose future and past event horizons intersect at the base event to which a
given pair is attached.

Let us call this demand the ``binary frame hypothesis''. It is expressed
quantum mechanically by the fact that ``achronal spin exists''. In other words,
``binary frames'' play a central role in physics only to the extent that
achronal spin does. The binary frame picture of spacetime and the existence
of achronal spin go hand in hand. They cannot be separated.

The binary aspect of a pair of accelerated frames becomes particularly
significant when interactions are present. In other words, it is of importance
that the domain of the two relativistic quantum systems include {\it both}
Rindler sectors. Much physical insight has been gained from the path breaking
considerations involving a multiparticle quantum (``quantized field'') system
interacting with an ``accelerated detector'' confined to a single Rindler
frame [2,16-20].  However, physics demands that an ``accelerated detector''
not discriminate against one of the Rindler components of the multiparticle
quantum system.  The ``accelerated detector'' should in fact consist of a {\it
pair} of detectors, one in each Rindler sector.  At a minimum one can always
recover the well known single detector results by tracing out the quantum
states of the other detector.

Using a pair of detectors is not a luxury.  Instead, it becomes a necessity if
one wishes to characterize the results of the interaction in terms of the {\it
conservation of total achronal spin} for the combined field-plus-``detector
pair'' quantum system.  Furthermore, using a pair of detectors becomes
unavoidable if one considers translations in spacetime.

The definition and conservation of achronal spin for the two interacting 
quantum systems is illustrated by the following simple and exactly soluble
example.

\subsection{6.2 Two Interacting Quantum Systems}

Consider two infinite quantum systems.  The first, the ``field system'', has
Klein Gordon quanta of rest mass $m$, while the second one, the ``detector pair
system'' has different quanta of rest mass $\mu$.  The domain of both systems
extends over all of spacetime, in particular over a given binary frame
consisting of the two Rindler frames, Eq.(1.3). Collisions between the $m$ and
the $\mu$ quanta are brought about by an interaction whose strength is
proportional to the coupling constant $e^2$. The wave equation for the two
systems are
$$\eqalignno{
(\bx - m^2)\psi&=e^2\phi  \cr
(\bx - \mu^2)\phi&=e^2\psi \quad& (6.1) \cr
}$$
Notice the simplicity of our assumption. It is expressed by the interaction 
Hamiltonian
$$\eqalignno{
H_{int}&= e^2\int _0^\infty 
                     (\psi^{\dag} \sigma_3 \phi+\phi^{\dag}\sigma_3 \psi) 
\xi d\xi  & \cr
&\equiv e^2 \int _0^\infty 
\lbrace 
[\psi^*_I~~ \psi^*_{II}]\sigma_3 
\left[ \matrix { \phi _I \cr  \phi _{II} \cr } \right] +
[ \phi^*_I~~ \phi^*_{II}]\sigma_3 
 \left[ \matrix {\psi _I \cr \psi _{II} \cr } \right] 
\rbrace \xi d\xi,  &(6.2)\cr
}$$
which is quadratic. The two wave equations can be decoupled by means of an 
orthogonal transformation, but we shall not have to do this.

Observable quantities and their dynamical evolution derive their importance
from their ability to carry the imprints of gravitation. The two key questions
are therefore
\hfil\break
(1) What are the observables?
\hfil\break
(2)What is their dynamical evolution as governed by the coupled equations
of motion, Eq.(6.1)?
\hfil\break
It is essential to answer these questions first in the absence of gravitation.
At another time, when gravitation is present the answer is different. This
difference is the imprint left by gravity.

\subsection{6.3 Achronal Spin and the Achronal Spin Four-Current}

To identify the appropriate observables, we incorporate the binary nature of
the pair of Rindler frames into a differential conservation law.
Considerations based on the inner product, Eq.(5.9), and the expectation values
of $\ora L $ imply that one introduce the {\it achronal spin four-current}
$$\eqalignno {
(\ora {J^\psi} )_\mu 
&= i \lbrace [\psi^*_I~~ \psi^*_{II}]\sigma_3 \ora L \left[ 
\matrix {\bigtriangledown _\mu \psi _I \cr \bigtriangledown _\mu \psi _{II} 
\cr }
\right] - 
[\bigtriangledown _\mu \psi^*_I~~ \bigtriangledown _\mu \psi^*_{II}]\sigma_3 
\ora L \left[ 
\matrix {\psi _I \cr \psi _{II} \cr }
\right] \rbrace \cr
&\equiv i \{ \psi^{\dag} ~ \sigma_3 \ora L 
\bigtriangledown _\mu \psi 
- \bigtriangledown _\mu \psi^{\dag}~ \sigma_3 \ora L \psi \}  & (6.3)\cr      
} $$
In the absence of interaction ($e^2=0$) its divergence vanishes. However, if
interactions are present, then with the help of Eqs.(6.1) the divergence is 
$$\eqalignno{
\bigtriangledown ^\mu (\ora {J^\psi} )_\mu
&= e^2i ( \psi^{\dag} \sigma_3 \ora L  \phi 
- \phi^{\dag} \sigma_3 \ora L \psi )  & (6.4)\cr      
} $$

The fourth component of the four-current is a density, that of achronal spin, 
as we presently shall see. The integral of this density
is the total amount of (vectorial) achronal spin relative to a given
binary frame. Using the inner product notation of Eq.(5.9) this vector is 
$$\eqalignno{
\ora {J^\psi} &=i \int _0 ^\infty 
( \psi^{\dag} \sigma_3 \ora L  {\partial \over \partial \tau}\psi 
- {\partial \over \partial \tau}\psi^{\dag} \sigma_3 \ora L \psi ) 
{d \xi \over \xi} \cr
&\equiv 2\langle \psi, \ora L \psi \rangle \quad . &(6.5) \cr
}$$
Within the framework of quantum mechanics both $\psi$ and  $\ora {J^\psi}$
are operators.
What compells us to identify the components $J^\psi _i~(i=1,2,3)$, of 
$\ora {J^\psi}$ as the achronal
spin components of the $\psi$ quantum system? Answer: The quantum mechanical
equal time (``$\tau$'') commutation relations
$$
\eqalignno{
[\dot \psi ^*_I (\xi,\tau),\psi _I(\xi ',\tau)]&= -i\xi\delta (\xi -\xi ') \cr
[\dot \psi ^*_I (\xi,\tau),\psi _{II}(\xi ',\tau)]&=0\cr
&~~\vdots\cr
&\hbox {etc.}  &\cr
}$$
of the $\psi$ system. They imply that
$$\eqalignno{
[J^\psi _1,J^\psi _2]&=-iJ^\psi _3\cr
[J^\psi _2,J^\psi _3]&=iJ^\psi _1\cr
[J^\psi _3,J^\psi _1]&=iJ^\psi _2  \quad .&(6.6)\cr}$$
These commutation relations of the quantum operators $J^\psi _i$ are those 
that characterize the $J^\psi _i$ as the achronal spin components, the 
generators of the achronal symmetry group $SU(1,1)$.

\subsection {6.4 Conservation of Achronal Spin}

The $\phi$ quantum system has its corresponding achronal spin $\ora {J^\phi}$
whose components also satisfy the $S(1,1)$ commutation relations
$$\eqalignno{
[J^\phi _1,J^\phi _2]&=-iJ^\phi _3\cr
[J^\phi _2,J^\phi _3]&=iJ^\phi _1\cr
[J^\phi _3,J^\phi _1]&=iJ^\phi _2  \quad .&(6.7)\cr}$$
The corresponding achronal spin four-current for the $\phi$ system is defined
in the same way as in Eq.(6.3), and its divergence is 
$$\eqalignno{
\bigtriangledown ^\mu (\ora {J^\phi} )_\mu
&= e^2i ( \phi^{\dag} \sigma_3 \ora L  \psi 
- \psi^{\dag} \sigma_3 \ora L \phi )\quad ,  & (6.8)\cr      
} $$
which does not vanish either. However, the divergence of total achronal spin
current, $(\ora {J^\psi} )_\mu +(\ora {J^\phi} )_\mu$ does vanish,
$$\eqalignno{
\bigtriangledown ^\mu [(\ora {J^\psi} )_\mu+(\ora {J^\phi} )_\mu]=0 &&(6.9)\cr
}$$

Thus one sees that in the presence of interaction ($e^2\neq 0$) none of the 
individual achronal spins, $\ora {J^\psi}$ and $\ora {J^\phi}$, are conserved.
However, the achronal spin of the total system,
$$\eqalignno{
\ora {J^{total}}=\ora {J^\psi} +\ora {J^\phi}, && (6.10)\cr
}$$
{\it is} conserved throughout the interaction process.

\subsection{6.5 Conserved Total Boost Energy and Total Charge}

In addition to the total achronal spin components, the binary frame observables
of the relativistic system include the total ``boost Hamiltonian''. Using the
the inner product notation of Eq.(5.9), this Hamiltonian is
$$\eqalignno{
H^{total}&= 2\langle \psi, i{\partial \over \partial \tau} \psi \rangle 
+2\langle \phi, i{\partial \over \partial \tau} \phi \rangle 
\quad . &(6.11a) \cr
}$$
for the two interacting systems $\psi$ and $\phi$. With the help of the wave 
Eqs.(6.1) and an integration this becomes the familiar sum
$$\eqalignno{
H^{total}=H^\psi +H^\phi +H_{int} &&(6.11b)\cr
}$$
of the $\psi$-system Hamiltonian
$$\eqalignno{
H^\psi &=\int_0 ^\infty \lbrace
{1\over {\xi ^2}} {\partial {\psi ^{\dag}}\over \partial \tau}\sigma_3
                 {\partial {\psi}\over \partial \tau}
+{\partial {\psi ^{\dag}}\over \partial \xi}\sigma_3
                 {\partial {\psi}\over \partial \xi}
+{\partial {\psi ^{\dag}}\over \partial x}\sigma_3
                 {\partial {\psi}\over \partial x}
+{\partial {\psi ^{\dag}}\over \partial y}\sigma_3
                 {\partial {\psi}\over \partial y}
+m^2 \psi^{\dag}\sigma_3 \psi \rbrace
\xi d\xi,\cr
&&(6.12)\cr
}$$
the analogous $\phi$-system Hamiltonian $H^\phi$, and the interaction 
Hamiltonian $H_{int}$, which is given by Eq.(6.2). Like the total achronal 
spin, $H^{total}$ is also conserved throughout the interaction process.

Besides the four conserved observables, Eqs.(6.10) and (5.1), there is also 
the conserved ``total charge'' operator Q
$$\eqalignno{
Q&=\langle \psi, \psi \rangle + \langle \phi, \phi \rangle ~~~~~~~~~~~~~~~~~~~~
\hbox{(total~charge)}
& (6.13)
}$$
for the two interacting Klein Gordon systems.

\section { 7. Spectral Decomposition}

For a multiparticle system the spectral decomposition of an observable
invariably highlights physical properties which would otherwise go unnoticed.
The achronal spin, boost energy and the charge of a free Klein-Gordon system
are no exceptions.  Consider the observables $\ora {J^\psi},~H^\psi $ and 
charge
$Q$ for a free $\psi$-system.  Taking note that its Heisenberg operator is given
by Eq.(5.18), one easily finds, with the help of Eqs.(5.9) and (6.11a), that
the spectral decompositions of Eq.(6.5), (6.12) and (6.13) are
$$\eqalignno{
J_1^\psi&={1\over 2}\int_{-\infty}^\infty
\lbrace (\tilde a^{\ast}_{\omega} \tilde b_{\omega}^{\ast} 
+\tilde a_{\omega} \tilde b_{\omega}) {\cosh \pi \omega \over \sinh \pi\omega}
+(\tilde a_{\omega} \tilde a_{\omega}^{\ast} 
+\tilde b_{\omega}^* \tilde b_{\omega}) {1 \over \sinh \pi\omega}
\rbrace d \omega \cr
J_2^\psi&={i\over 2}\int_{-\infty}^\infty
(\tilde a_{\omega} \tilde b_{\omega} -\tilde a^*_{\omega} \tilde b^*_{\omega})
d \omega &(7.1)\cr
J_3^\psi&={1\over 2}\int_{-\infty}^\infty
\lbrace (\tilde a^{\ast}_{\omega} \tilde b_{\omega}^{\ast} 
+\tilde a_{\omega} \tilde b_{\omega}) {1\over \sinh \pi\omega}
+(\tilde a_{\omega} \tilde a_{\omega}^{\ast} 
+\tilde b_{\omega}^* \tilde b_{\omega}) {\cosh \pi \omega\over \sinh \pi\omega}
\rbrace d \omega \cr
H^\psi&=\int_{-\infty}^\infty \omega
(\tilde a_{\omega} \tilde a^*_{\omega} -\tilde b^*_{\omega} \tilde b_{\omega})
d \omega &(7.2)\cr
Q&={1\over2}\int_{-\infty}^\infty 
(\tilde a^*_{\omega} \tilde a_{\omega} -\tilde b_{\omega} \tilde b^*_{\omega})
d \omega \quad .&(7.3)\cr
}$$
One readily verifies the familiar $SU(1,1)$ commutation relations
$$\eqalignno{
[J^\psi _1,J^\psi _2]&=-iJ^\psi _3\cr
[J^\psi _2,J^\psi _3]&=iJ^\psi _1\cr
[J^\psi _3,J^\psi _1]&=iJ^\psi _2  \quad .&(7.4)\cr
\noalign {\hbox {and~also}}
[H^\psi,\ora {J^\psi}]&=0 &(7.5)\cr
     [Q,\ora {J^\psi}]&=0 \quad .&\cr
}$$
Thus, as expected, achronal spin is invariant under boost.

Note that an equivalent set of achronal spin operators would have been
$$\eqalignno{
J_1^{'\psi}&={1\over 2}\int_{-\infty}^\infty
(\tilde a^{\ast}_{\omega} \tilde b_{\omega}^{\ast} 
+\tilde a_{\omega} \tilde b_{\omega}) 
d \omega \cr
J_2^{'\psi}&={i\over 2}\int_{-\infty}^\infty
(\tilde a_{\omega} \tilde b_{\omega} -\tilde a^*_{\omega} \tilde b^*_{\omega})
d \omega &(7.6)\cr
J_3^{'\psi}&={1\over 2}\int_{-\infty}^\infty
(\tilde a_{\omega} \tilde a_{\omega}^{\ast} 
+\tilde b_{\omega}^* \tilde b_{\omega})  d \omega .\cr
}$$

Both the original $J_i^\psi$'s and the newly defined $J_i^{'\psi}$'s obey the
same commutation relations. If one pictures the spectral components of
$J_1^{'\psi}$, $J_2^{'\psi}$, and $J_3^{'\psi}$ as three orthogonal basis
vectors in a three dimensional Lorentz space, then the spectral components of
$J_1^{\psi}$, $J_2^{\psi}$, and $J_3^{\psi}$ form an alternate
Lorentz-orthogonal basis with $J_3^\psi$ pointing along the direction of the
Planckian vector, Eq.(5.12). The Lorentz rotation which relates the two sets of
basis vectors was induced by the spinor transformation which relates Eq.(5.15)
to Eq.(5.18).

\section{8. Why Achronal Spin?}

To answer this question let us first compare (i) a system of cavity modes in
thermalequilibrium and (ii) a single-charge quantum system viewed relative to
spacetime-translated paired accelerated frames. 

Both systems carry the Planckian power and fluctuation spectra as their
characteristic signature. However considerably more important is their
difference. In one case these spectra express, via the moments of a thermal
density matrix, the statistical equilibrium of the infinity system. In the
second case, they express,via the expectation values relative to pure quantum
states, the spacetime translation invariance of the single-charge system.

What does this mixed state versus pure state distinction signify? For the
cavity mode system the significance is well known: the two spectra, Planckian
and fluctuation, imply the existence of photons within the statistical
thermodynamic setting. This we have learned from Planck and Einstein. 

Is there a corresponding significance for the single-charge system? Let us
assume that the spectra of quantum mechanical expectation values given by
Eq.(5.22) or (5.23) lend themselves to being measured by an experiment. If
carried out, such an experiment would imply the existence of achronal spin, a
quantum mechanical expression of the acceleration-induced partitioning of
spacetime into pairs of adjacent but causally disconnected
(i.e. ``achronally'' related) Rindler sectors.

Achronal spin is not limited to single-charge quantum systems. Its extension
to infinite (e.g. multimode) quantum systems leads to the achronal spin
density-current, which satisfies a differential conservation law. As a consequence {\it total achronal spin is neither created nor destroyed.} This makes
achronal spin a macroscopic property of matter.

The second reason for achronal spin is geometrical.
Different spacetime frameworks demand different ways of characterizing
the dynamics of physical systems, for example, the two interacting quantum 
systems described in Section 6.
\hfil\break
(1) Relative to an inertial frame use energy, momentum, and charge.
\hfil\break
(2) Relative to a binary frame use achronal spin, boost energy, and charge.

Thus the use of energy-momentum as a property of matter is predicated on
locally translation invariant inertial (``free fall'') frames. In other words,
to each event of spacetime there is attached what mathematicians call a
tangent space. By contrast, the use of ``boost energy'' and achronal spin as a
property of matter is predicated on paired accelerated frames, one paired frame
attached to each event of spacetime.

The geometrical difference, inertial frames vs. paired
accelerated frames, goes hand in hand with with the physical difference,
energy and momentum vs. achronal spin and boost energy.

Which do we choose, and why?  The choice is between two geometrical views of
flat Minkowski spacetime.  Its invariance group is
the Poincare group, the (semi) direct product of translations and Lorentz
rotations around an event.

The first choice is implemented by trivializing translation invariance, i.e.
by expressing quantum mechanical processes in terms of inertial, i.e.
translation eigenfunctions (``plane waves'').  This is done in all the text
books.  Lorentz invariance is imposed as an afterthought on the tangent space
at each event.  This imposition is done by endowing that tangent space with a
Lorentz metric and by requiring Lorentz invariance, i.e. that all physical(ly
measurable) properties be expressible in terms of the scalars, vectors, spinors
etc. under Lorentz transformations.

The second choice is implemented by trivializing Lorentz invariance, i.e. by
expressing quantum mechanical processes in terms of boost eigen functions
(for a 2-D Rindler frame) or their generalization (for a 4-D spherical Rindler
frame as in Figure 1).  Physical properties in different pairs of accelerated
frames are then related by the appropriate translation transformation.

In brief, the difference between the two choices consists of interchanging the
role that the translation group and the Lorentz play in physical processes.

The choice of groups is, however, not merely a matter of taste.  In a pair of
accelerated frames the boost eigenstate description of a particle implies a 
qualitatively new feature, the particle's achronal spin.  An inertial frame
with its translation eigenstates can not make any such claim.

\section{References}

\medskip

\item{[1]}
P. C. W. Davies, J. Phys. A: Math. Gen. 8, 365 (1975).

\item{[2]}
W. G. Unruh, Phys. Rev. D 15, 870 (1976).

\item{[3]}
P. Candelas and D. Deutsch, Proc. R. Soc. London A354, 79 (1977)

\item{[4]}
P. Candelas and D. Deutsch, Proc. R. Soc. London A362, 251 (1978)

\item{[5]}
M. Soffel, B. Mueller, W. Greiner, Phys. Rev. D22, 1935 (1980)

\item{[6]}
W. G. Unruh and N. Weiss, Phys. Rev. D29, 1656 (1984)

\item{[7]}
J. A. Wheeler, {\it A Journey into Gravity and Spacetime}
(Scientific American Library, New York, 1990)

\item{[8]}
C.W. Misner, K.S. Thorne, and J.A. Wheeler, {\it Gravitation}
(W.H. Freeman and Co., San Francisco, 1973) ch. 15

\item{[9]}
U.H. Gerlach, Phys. Rev. D28, 761 (1983)

\item{[10]}
The $\theta ,\phi ,~ {\rm  and }~\tau$ coordinates are what 
in Euclidean space correspond to the Euler parameters on a three-sphere. They
are not to be confused with the $\theta ,\phi ,~ {\rm  and }~\tau$ coordinates
of Eq.(4.1), which are what in Euclidean space corresponds to the usual 
spherical coordinates of a three-sphere. The coordinate transformation Eq.(4.4)
is a lower dimensional adoptation of the one Rindler [W. Rindler, Found. of 
Physics 15, 545 (1985)] used to identify the world line dependent event
horizons on the De Sitter hyperboloid.

\item{[11]}
C.W. Misner, K.S. Thorne, and J.A. Wheeler, {\it Gravitation}
(W.H. Freeman and Co., San Francisco, 1973) p880

\item{[12]}
Obtained by inverting Eqs.(4.2) in U.H. Gerlach, Phys.Rev. D 38, 514 (1988),
or in gr-qc/9910097 .

\item{[13]}
Integration over $x$ and $y$ is understood, but suppressed throughout this 
article.

\item{[14]}
Reference to the modes $\exp (ik_x x)$ and $\exp (ik_y y)$ as well as the 
corresponding mode integrations are understood, but suppressed throughout 
this article.

\item{[15]}
The arguement leading to the angle of inclination, Eq.(5.24), from the 
precession axis (5.22) holds only for a superposition of two plane wave
states. It does not hold for a general superposition of plain wave states, as 
claimed in the original version of this article published in the MG7
Proceedings.

\item{[16]}
B. DeWitt in {\it General Relativity: An Einstein centenary survey,} edited
by S.W. Hawking and W. Israel (Cambridge University Press, New York, 1979)
Ch. 14.2.2

\item{[17]}
N.D. Birrell and P.C.W. Davies, {\it Quantum fields in curved space}
(Cambridge University Press, New York, 1984) Ch. 3.3

\item{[18]}
W.G. Unruh and R.M. Wald, Phys. Rev. D29, 1047 (1984)

\item{[19]}
D.J. Raine, D.W. Sciama, and P.G. Grove, Proc. Roy. Soc.A435, 205 (1991);
P.G. Grove, Class. Quantum Grav. 3, 801 (1986)

\item{[20]}
W.G. Unruh, Phys. Rev. D46, 3271 (1992)

\end